\documentclass[conference,10pt,a4paper]{IEEEtran}
\usepackage{url,amsmath,cite,caption}
\ifx\pdfoutput\undefined
  \usepackage[pdftex]{graphicx}
\else
  \usepackage{graphicx}
\fi

\usepackage{array}

\begin{document}

%\title{A simple statistical analysis approach for Intrusion Detection System}% for Coordinated Attack}
%Judul ini juga dikomentari pak, katanya: "Title more suitable as: "Intrusion Detection Technique based on Statistical Analysis"
\title{A simple statistical analysis approach for Intrusion Detection System}

\author{\authorblockN{A.A. Waskita\authorrefmark{1}\authorrefmark{2}, 
H. Suhartanto\authorrefmark{2}, P.D. Persadha\authorrefmark{3}\authorrefmark{2},
L.T. Handoko\authorrefmark{4}\authorrefmark{5}\authorrefmark{6}}
\authorblockA{\authorrefmark{1}Center for Development of Nuclear
Informatics, National Nuclear Energy Agency, \\
Kawasan Puspiptek Serpong, Tangerang 15310, Indonesia\\
Email : adhyaksa@batan.go.id}
\authorblockA{\authorrefmark{2}Faculty of Computer Science, University of
Indonesia, \\
Kampus UI Depok, Depok 16424, Indonesia\\
Email : heru@cs.ui.ac.id}
\authorblockA{\authorrefmark{3}Sub Directorate of Ciper Security Technique,
National Crypto Agency, \\
Jl. Harsono RM No.70 Ragunan Ps.Minggu Jakarta Selatan, Indonesia-12550\\
Email: pratama.dahlian@lemsaneg.go.id}
\authorblockA{\authorrefmark{4}Group for Theoretical and Computational Physics,
Research Center for Physics, Indonesian Institute of Sciences, \\
Kawasan Puspiptek Serpong, Tangerang 15310, Indonesia\\
Email: laksana.tri.handoko@lipi.go.id}
\authorblockA{\authorrefmark{5}Research Center for Informatics, Indonesian Institute of Sciences\\
  Kompleks LIPI Cisitu, Jl. Cisitu 21/154D, Bandung 40135, Indonesia}
\authorblockA{\authorrefmark{6}Department of Physics, University of Indonesia,\\
Kampus UI Depok, Depok 16424, Indonesia\\
Email: handoko@fisika.ui.ac.id}
}

\maketitle

\begin{abstract}
A novel approach to analyze statistically the network traffic raw data is proposed. The huge amount of  raw data of actual network traffic from the Intrusion Detection System is analyzed to determine if a traffic is a normal or harmful one. Using the active ports in each host in a network as sensors, the system continuously monitors the incoming packets, and generates its average behaviors at different time scales including its variances. The average region of behaviors at certain time scale is then being used as the baseline of normal traffic. Deploying the exhaustive search based decission system, the system detects the incoming threats to the whole network under supervision.
\end{abstract}


\section{Introduction}
\label{sec:intro}

Securing a system from harmful disruptions, either from the system itself or the external attacks is an essential problem since the initial internet age. Usually, a system, especially those are characterized as a safety critical system employs a number of sensors to detect any influential parameters \cite{SciTopics, aircraftSafety, CarvalhoNormal, ftaOilTank, uncertainBondGraph, BjornesethMaritime}. The acquired value of parameters are than analyzed with particular techniques to determine a system under operation is always in a safe condition \cite{jamilAffandi,timeseriesNN,recurrentNN,CarlosPSO}.

In case of securing computer network system, beside of keeping the system is at its optimal condition to serve the users, keeping the sensitive data from inappropriate or unauthorized users is also an important issue. In securing the system from any kind of intrusions some sensor like components play an important role \cite{simmons2009avoidit}. Instead of obtaining physical parameters, the sensors have a role in monitoring traffic accross the network. 

Traffic monitoring is in general the responsibility of Intrusion Detection System (IDS). It captures packets at particular time frame, namely every miliseconds. The whole historical data is analyzed carefully to perform a kind of decision support system to detect any inappropriate access at nearly real time manner. From those captured packets, IDS extracts more detail information such as packet size, the origin of IP address, the attacked port number and also its packet type like ICMP, TCP and UDP as well \cite{AgarwalHybrid,snort}. In analyzing the traffic data, IDS might look at the content of the packet to determine whether it contains any malicious code. Otherwise, it make use of any deviations from normal behaviors and profiles. The normal behaviors represent the normal or expected behaviors derived from previous regular activities, network connections, hosts or users over certain period of time. The first method in this kind is categorised as the signature-based IDS, while the last one is known as the anomaly-based IDS \cite{idsReview}. The signature-based IDS has a lower false alarm than the anomaly-based IDS. It uses a set of rules known as signatures to identify an intrusion packet \cite{AdaBoost}. However, it identifies the unknown intrusions inefficiently. On the other hand, by identifying deviation from the normal behaviors, the anomaly-based IDS is able efficiently to detect a novel intrusion. However, it produces higher rate of false alarm.

In the matter of traffic data analysis, some research works have previously been conducted in various approaches for various target. For instance, the work in \cite{AdaBoost,kMeans} proposed a method to improve the detection rate. While the reference \cite{ashokSVM,MultipleKernel} tried to reduce the unuseful features, or reference \cite{OutlierKe} deployed the adaptive IDS. However, all of them were applied to the KDD'99 data set which is not recommended to be used for testing the algorithms \cite{KDnuggets,Mahbod}. Another researches addressed some tools for network engineer to use easily the IDS \cite{Association}, and to smooth the boundary of each classification based on the artificial intelligence (AI) \cite{fuzzyAssociation}. For a special case where the computer network is operated in a safety critical environment, the system should be able to generate detection rate with high accuracy if not absolutely correct. The AI based approaches, by default, have its limitation coming from the reduced search space to speed up the computing time.

In this paper, we propose a simple statistical based analysis to analyze the traffic accross the ports in every host in a network. Such ports are treated as a number of sensors in a host running certain applications. A number of hosts then construct a computer network system receiving the attacks. Instead of capturing physical parameters as done in our previous work \cite{arya}, the IDS mines the whole network traffic data, and associate them to the appropriate port as a cumulative event of incoming or outgoing traffic at particular time scale. A cumulative event of traffic in that time scale, including its variance, is subsequently assigned to each port. All of them are then treated in a similar manner as the physical sensors in \cite{arya}. A number of incoming and outgoing packets from a certain port is a common parameter used in analyzing network traffic data \cite{gabra2012data}. Besides using those similar parameters \cite{VerwoerdIDSApproach} or methods \cite{vaarandi,dzhao}, the present approach  also considers the interactions among different ports to determine how far the deviation of a current status from the known patterns. It is arqued that concerning the port interactions exhaustively should increase the accuracy of detection rate. The reason is simply because of no search spaces are reduced.

This paper is organized as follow. First, in Sec. \ref{sec:intro} the backgroud is presented. The proposed model is then presented in Sec. \ref{sec:model} where the basic architecture and a detail description of the system is explained. The paper is ended with a summary.

\section{The Model}
\label{sec:model}

The model is based on our previous work \cite{arya}. In order to treat various sensors with different specifications (the operation range, etc), one should perform a kind of normalization procedure. This would guarantee the acquired physical measurements from any sensors are always at the same range, namely between $0$ to $1$. Though, in real life there is no any physical parameter involved in a computer network system. Further,  a number of incoming and outgoing packets to a certain port is treated as the basic numerical information like the acquired physical measurements. By averaging the whole data in certain period of time, either minute, hour, day or month, its variance or maximum and minimum values can be obtained. These information are analogous to the range of physical sensors. Therefore, the problem is now turned into the sensor network as done in 
 \cite{arya}.
 
\begin{figure}[t!]
 \centering
 \includegraphics[scale=.45]{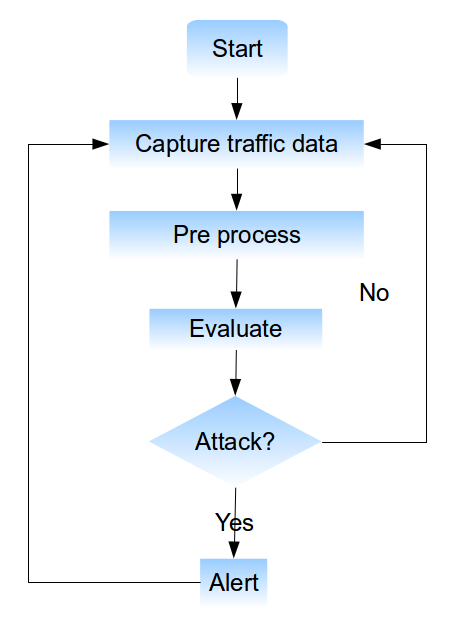}
\caption{The flowchart of capturing the network traffic data and its analysis.}
\label{fig:flowchart}
\end{figure}

Most of AI based approaches are classified as the supervised method, where a number of training data must initially be provided. In constrast with AI, this approach should be classified as the supervised and unsupervised ones. There is no static training data since the data is captured continuously to the network traffic data is performed continuously as shown in Fig. \ref{fig:flowchart}.

\begin{figure}[b!]
 \centering
 \includegraphics[scale=.42]{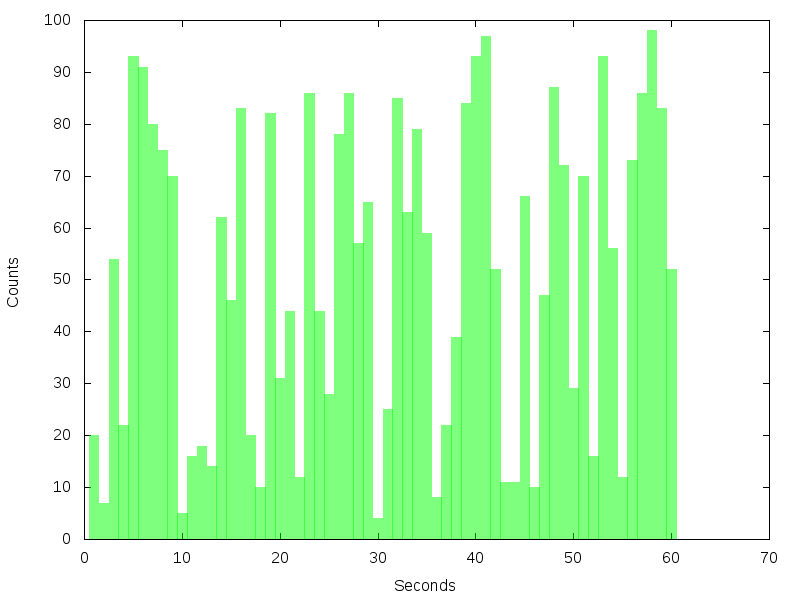}
\caption{The illustration of graph generated from the incoming packets to a certain port in every seconds.}
\label{fig:seconds}
\end{figure}

The system constinuously captures the traffic step by step from very short time period to relatively longer ones. 
%Perbaikan di sini awalnya karena ada komentar "Another Figure to resemble normalization effect of Figure 2 should be included too". Lalu saya terpikir untuk menjelaskannya lebih detil, bahwa saat datang dengan skala milli second dan setiap second ditabulasi. Hasil tabulasi itu yang menghasilkan data rata-rata untuk skala menit.
For instance,  the shortest time period is at the order of second as illustrated in Fig. \ref{fig:seconds} can be obtained from the accumulated number of incoming or outgoing packet every millisecond. The average and deviation are obtained directly from the raw data. In the case of a computer network receives much lower traffic, the second time basis may be useless. In this case it might be plausible to use longer time scale such as minute.

The  captured data will then be pre-processed by averaging over the time frame. The average of incoming or outgoing packet to / from a certain port at a period of time $\Delta t$ can be written as,
\begin{equation}
  N_{min}^{i} = \frac{n_{1} + n_{2} + \ldots + n_{t}}{t} \; ,
  \label{eq:Nminute}
\end{equation}
where $N_{min}^{i}$ is an average number of incoming or outgoing packets during the time period in the minute basis for the $i^\mathrm{th}$ minute. 
%Maaf pak, maksud saya, ini untuk menotasikan N sebagai nilai rata-rata untuk skala waktu menit (min) untuk menit ke-i. Nanti, klo rata-rata jumlah paket tiap jam, jadinya N_hour. Itu gimana nulisnya pak?
For example, there would be $60$ average number of packets in every second.

\begin{figure}[t!]
 \centering
 \includegraphics[scale=.32]{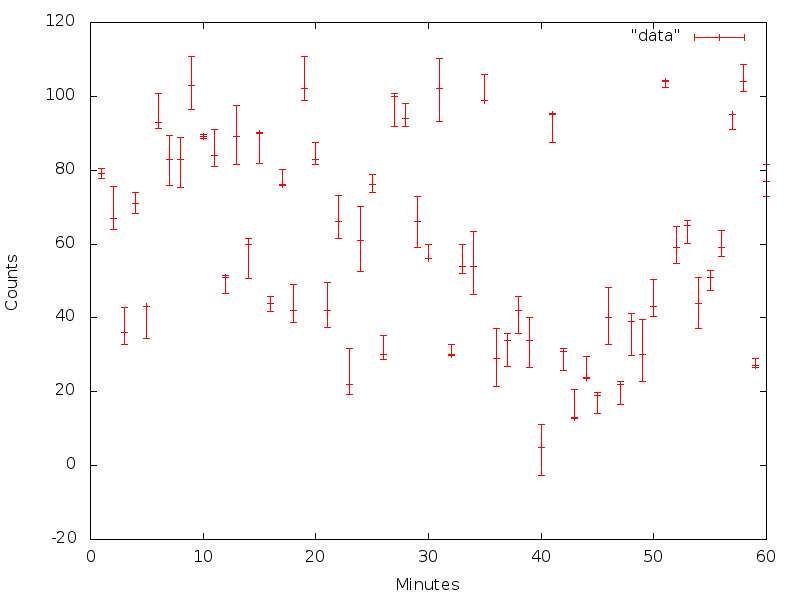}
\caption{The illustration of graph generated from the incoming packets to a certain port in every minutes.}
\label{fig:minutes}
\end{figure}

From the obtained raw data, one can plot some figures across the time scale, either minute, hour, day or even month againts the number of packets. An example is given in Fig. \ref{fig:minutes} shows the packet fluctuations every minute in an hour of period. One can then simply conclude that an average of number of packet at a certain time scale could be used as the baseline region within its errors if an intrusion traffic falls outside the region. Because the variances in every averaged data should have similar characteristic with the previous normal accumulated packets. Furthermore, normalizing all of them into $0 \sim 1$ will translate exactly the current problem to the sensor network problem in our previous work. 

From Fig. \ref{fig:minutes}, it is clear that at every minute scale, there are maximum ($x_{max}$) and minimum ($x_{min}$) values. Those values should be considered as the operation ranges of a physical sensor in \cite{arya}. Thus, the same formula for normalizing the values,
\begin{equation}
  x_i = f \, \left( x^\prime_i - x_\mathrm{min} \right) \; ,
  \label{eq:normalizing}
\end{equation}
can be adopted as well. Following the same procedure, the value of normalization factor $f$ can be determined by,
\begin{equation}
  f = \frac{1}{\left| x_\mathrm{max} - x_\mathrm{min} \right|} \; .
  \label{eq.normFactor} 
\end{equation}

To generate similar figures for hourly time scale, one can subsequently average the mean values of the preceeding numbers of packet in the minute time scale. For further time scales (daily, monthly, annually and so on) can be calculated in the same manner. However, it should be noted that calculating the average values for different time scale must be done in a subsequent order. It means that, in order to calculate the average value of a monthly time scale, one should perform the preceeding daily time scale. It should also be generated from the daily time scale and so forth.

After successfully generating the data order by order, one should proceed with the normalization procedure. This is crucial to enable proper evaluation among different ports at the same level. The evaluation process is the same as \cite{arya}. However, beside using simple general static threshold value, in the present case the value will dinamically be calculated along the operation time. The main reason is the so-called 'sensors' in the present case are not the physical ones with pre-defined upper and lower ranges. In contrary, its upper and lower ranges are volatile depending on the time period of data acquisition, since it simply counts the packets. This point is the main difference with the previous work which assuming a system of network 'physical' sensors.
 
\begin{figure}[t!]
 \centering
 \includegraphics[scale=.32]{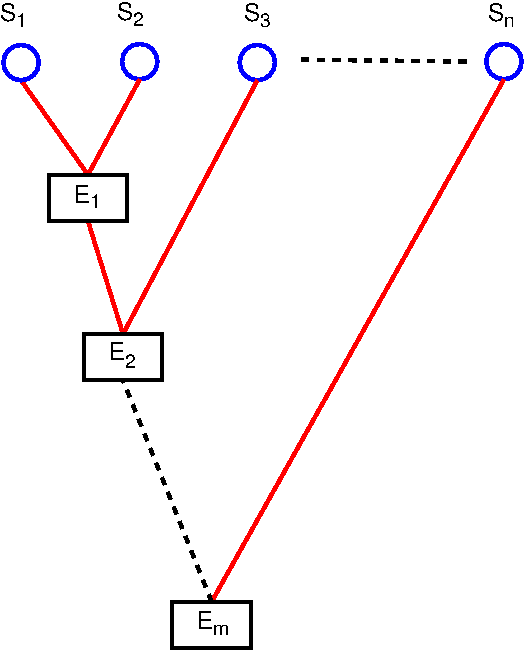}
\caption{The illustration of possible active ports relationship.}
\label{fig:relation}
\end{figure}

Concerning the interactions among the ports, one can still relate the active ports each other. The relationship among active ports can be similar to one illustrated in Fig. \ref{fig:relation}. However, in the present case one should consider the strength of interactions as the representation of the degree of similarness in terms of protocols being used to access the ports. Consequently, the relationship is not always being able to be modeled linearly.

\section{Discussion}

As previously described, the maximum and minimum value pairs as depicteg in Fig. \ref{fig:minutes} are  analogous to an operation range of physical sensors. In the case of computer network security, the values can be used as an indication of incoming attacks. If the accumulated packets on a certain port at a certain period lies outside the maximum and minimum value pairs, it could be considered as an attack. However, a number of maximum and minimum value pairs which is a tool to indicates the occurence of attacks should be generated from averaging the accumulated packets in various time scale and period of data collection. Therefore the problem is turned into how to define correct time scale and period of data collection to obtain high detection rate of the IDS.

Statistically, averaging a number of data collected in a longer period could result much less accurate  description of the object. The reason is straightforward, because it leads to wider ranges of maximum and minimum value pairs which is classified as a 'normal' region. On the other hand, it is in general advisable to set up shorter period of data colection to determine the normal region. Of course performing all incoming traffics with this scheme could be extravagant. Therefore, one should perform step by step comparison with the allowed regions generated from long to shorter period of time.

The above detection scheme requires more computating power. This is due to the requirement to dynamically recalculated the maximum and minimum value pairs of different time scale and period of data collection. As illustrated in Fig. \ref{fig:relation}, the sensors like $S_{1}$ are represented by a normalized value as calculated in Eq. \ref{eq.normFactor} which requires additional computating power. Hence, one may utilize the distribution job scheme into different machine as suggested in \cite{arya}.

Moreover, dynamically recalculating the above parameters in many different time scales and periods of data collection, require an efficient data store mechanism as well. Here, efficient means that the system only stores the minimum amount of a pre-processed data, and is able to retrieve a proper data to be calculated dynamically over different time scale and period of data collection.

\section{Summary}
\label{sec:summary}

A new kind of modeling the IDS using the straightforward statistical approach has been presented. It is argued that the basic statistical approach might still be appropriate if no search space is reduced. This is realized by deploying the exhaustive search method which has recently been developed for a system of network sensor. 

However, more comprehensive simulation with appropriate data set, distribution scheme of computational power with efficient data storing system is still undertaken, and will be presented soon elsewhere. 

\section*{Acknowledgements}

AAW and PDP greatly thank the Indonesian Ministry of Research and Technology for financial support during the work. LTH  is funded by the Riset Kompetitif LIPI in fiscal year 2013 under Contract no.  11.04/SK/KPPI/II/2013.

\bibliographystyle{IEEEtran}
\bibliography{icspc}
\end{document}